\documentclass{PoS}

\title{The most distant radio quasars at the highest resolution}

\ShortTitle{The most distant radio quasars at the highest resolution}

\author{\speaker{S\'andor Frey}$^{,a,d}$, Zsolt Paragi$^{b,d}$, Leonid I. Gurvits$^{b}$, Krisztina \'E. Gab\'anyi$^{a,d}$, D\'avid Cseh$^{c}$\thanks{The EVN is a joint facility of European, Chinese, South African, and other radio astronomy institutes funded by their national research councils. This work was supported by the European Community's Seventh Framework Programme, Advanced Radio Astronomy in Europe, grant agreement no.\ 227290, the European Community's Seventh Framework Programme (FP7/2007-2013) under grant agreement no.\ ITN 215212 ``Black Hole Universe'', and the Hungarian Scientific Research Fund (OTKA, grant no.\ K72515).}\\
\llap{$^a$} F\"OMI Satellite Geodetic Observatory, P.O. Box 585, H-1592 Budapest, Hungary\\ 
\llap{$^b$} Joint Institute for VLBI in Europe, Postbus 2, 7990 AA Dwingeloo, The Netherlands\\
\llap{$^c$} Laboratoire Astrophysique des Interactions Multi-echelles (UMR 7158), CEA/DSM-CNRS-Universit\'e Paris Diderot, CEA Saclay, F-91191 Gif sur Yvette, France\\
\llap{$^d$} MTA Research Group for Physical Geodesy and Geodynamics, P.O. Box 91, H-1521 Budapest, Hungary\\
E-mail: \email{frey@sgo.fomi.hu}, \email{zparagi@jive.nl}, \email{lgurvits@jive.nl}, \email{gabanyik@sgo.fomi.hu}, \email{david.cseh@cea.fr}} 

\abstract{There are about 50 quasars known at redshifts $z>5.7$ to date. Only three of them are detected in the radio (J0836+0054, $z=5.77$; J1427+3312, $z=6.12$; J1429+5447, $z=6.21$). The highest-redshift quasars are in the forefront of current astrophysical and cosmological research since they provide important constraints on the growth of the earliest supermassive black holes in the Universe, and on the physical conditions in their environment. These sources are indeed associated with active galactic nuclei as revealed by high-resolution Very Long Baseline Interferometry (VLBI) observations. It is still unclear whether the physical properties of the few $z$$\sim$6 radio quasars are in general similar to those of their lower-redshift cousins. In the case of J1427+3312, the 100-pc scale double morphology suggests a young radio source. Here we report on the recent European VLBI Network (EVN) imaging observations of J1429+5447, the most distant radio quasar known. Based on its milli-arcsecond-scale structural and spectral properties, this quasar is similar to J0836+0054 and J1427+3312. This raises the question if the compact steep-spectrum radio emission is a ``universal'' feature of the most distant radio quasars. It could only be answered after many more objects at $z>6$ are discovered and studied.}

\FullConference{10th European VLBI Network Symposium and EVN Users
Meeting: VLBI and the new generation of radio arrays \\
		 September 20-24, 2010\\
		 Manchester, UK}

\begin{document}

\section{Introduction}

Observations of quasars at the highest redshifts can constrain models of the early cosmological evolution of active galactic nuclei (AGNs) and the growth of their central supermassive (up to $\sim$$10^9$~$M_{\odot}$) black holes. Currently the object CFHQS~J0210$-$0456 holds the redshift record among the quasars with $z$=6.44 \cite{Will10a}. There are about 50 quasars known at redshifts $z>5.7$ to date. Interestingly, despite the growing number of known $z$$\sim$6 quasars, the maximum redshift did not increase much over the last decade or so (e.g. SDSS~J1148+5251, $z$=6.43 \cite{Fan03}). It remains to be seen whether it is a selection effect due to current limitations of the high-redshift identification techniques, or the first quasars in the Universe indeed started to ``turn on'' at around this cosmological epoch. Intriguingly, most of the observed properties of the highest-redshift quasars are remarkably similar to those of the lower-redshift objects. Only recently there were found a couple of $z$$\sim$6 quasars which do not show infrared emission originating from hot dust, and the amount of hot dust may increase in parallel with the growth of the central black hole \cite{Jian10}. This suggests that at least some of the most distant quasars known are not completely evolved objects. 

Only three of the $z>5.7$ quasars (J0836+0054, $z=5.77$; J1427+3312, $z=6.12$; J1429+5447, $z=6.21$) are detected in the radio. Which makes them particularly valuable is that the ultimate evidence for AGN jets can be provided by high-resolution Very Long Baseline Interferometry (VLBI) observations. The radio structure of J0836+0054 on $\sim$10 milli-arcsecond (mas) angular scale (\cite{Frey03,Frey05}) is characterised by a single compact but somewhat resolved component, with steep radio spectrum ($\alpha$=$-0.8$) between the observed frequencies of 1.6 and 5~GHz. (The power-law spectral index $\alpha$ is defined as $S\propto\nu^{\alpha}$, where $S$ is the flux density and $\nu$ the frequency.) The VLBI images of the first $z>6$ radio quasar, J1427+3312 (\cite{Frey08,Momj08}) revealed a prominent double structure. The two resolved components are separated by $\sim$28~mas ($\sim$160~pc). (To calculate linear sizes and luminosities, we assume a flat cosmological model with $H_{\rm{0}}=70$~km~s$^{-1}$~Mpc$^{-1}$,  $\Omega_{\rm m}=0.3$, and $\Omega_{\Lambda}=0.7$.) The structure similar to that of the compact symmetric objects (CSOs) is one of the indications of the youthfulness of J1427+3312. The brighter component also detected at 5~GHz has a steep radio spectrum ($\alpha$=$-0.6$).

A recent census of VLBI-imaged radio quasars at $z>4.5$, and European VLBI Network (EVN) imaging of five new sources at $4.5 < z < 5$ was made by \cite{Frey11}. The slightly resolved mas- and 10-mas-scale radio structures, the measured moderate brightness temperatures ($\sim$10$^8$$-$10$^9$~K), and the steep spectra in the majority of the cases suggest that the sample of compact radio sources at $z>4.5$ is dominated by objects that do not resemble blazars that are characterised by highly Doppler-boosted, compact, flat-spectrum radio emission.

According to the model of \cite{Falc04}, the high-redshift steep-spectrum objects may represent gigahertz-peaked-spectrum (GPS) sources at early cosmological epochs. The first generation of supermassive black holes could have powerful jets that developed hot spots well inside their forming host galaxy, on linear scales of 0.1$-$10~kpc. Taking the relation between the source size and the turnover frequency observed for GPS sources into account, and for hypothetical sources matching the luminosity and spectral index of our ``typical'' quasars at $z$$\sim$5 or higher, the angular size of the smallest ($\sim$$100$~pc) of these early radio-jet objects would be in the order of 10~mas, and the observed turnover frequency in their radio spectra would be around 500~MHz \cite{Falc04}.

The quasar CFHQS~J142952+544717 (J1429+5447 in short) was discovered in the Canada-France High-z Quasar Survey (CFHQS; \cite{Will10b}). With the spectroscopic redshift of $z$=6.21, it is the most distant radio-loud quasar known to date. The object appears in the Very Large Array (VLA) Faint Images of the Radio Sky at Twenty-centimeters (FIRST) survey \cite{Whit97} list as an unresolved ($<5^{\prime\prime}$) radio source with an integral flux density of $S$=2.95~mJy at 1.4~GHz. We observed J1429+5447 with the EVN at 1.6 and 5~GHz, to compare its high-resolution radio structure and spectral properties with those of the two other $z$$\sim$6 quasars already known. Here we report on the preliminary results of our analysis.

\section{EVN observations of J1429+5447}

The observations took place on 2010 May 27 (5~GHz frequency; experiment EF022a) and on 2010 June 8 (1.6~GHz; EF022b). At a recording rate of 1024~Mbit~s$^{-1}$, eleven antennas of the EVN participated in the experiment at 5 GHz: Effelsberg (Germany), Jodrell Bank Lovell \& Mk2 telescopes (UK), Medicina (Italy), Toru\'n (Poland), Onsala (Sweden), Sheshan, Nanshan (P.R. China), Badary, Zelenchukskaya (Russia), and the phased array of the Westerbork Synthesis Radio Telescope (WSRT, The Netherlands). All but the Jodrell Bank Mk2 telescope participated in the 1.6-GHz experiment as well. Both experiments lasted for 6 h. Eight intermediate frequency channels (IFs) were used in both left and right circular polarisations. The total bandwidth was 128~MHz per polarisation. The correlation of the recorded VLBI data took place at the EVN Data Processor at the Joint Institute for VLBI in Europe (JIVE), Dwingeloo, the Netherlands.

The weak target source, J1429+5447, was observed in phase-reference mode to increase the coherent integration time spent on the source and thus to improve the sensitivity of the observations. Phase-referencing involves regularly interleaving observations between the target source and a nearby bright and compact reference source. The phase-reference calibrator J1429+5406 is separated from the target by 0.69$^{\rm o}$ in the sky. The target--reference cycles of $\sim$5.5~min allowed us to spend $\sim$3.5~min on the target source in each cycle, providing almost 3 h total integration time on J1429+5447. Phase-referencing also allows us to determine the accurate relative position of the target source with respect to the well-known position of the reference source. 

\begin{figure}[!h]
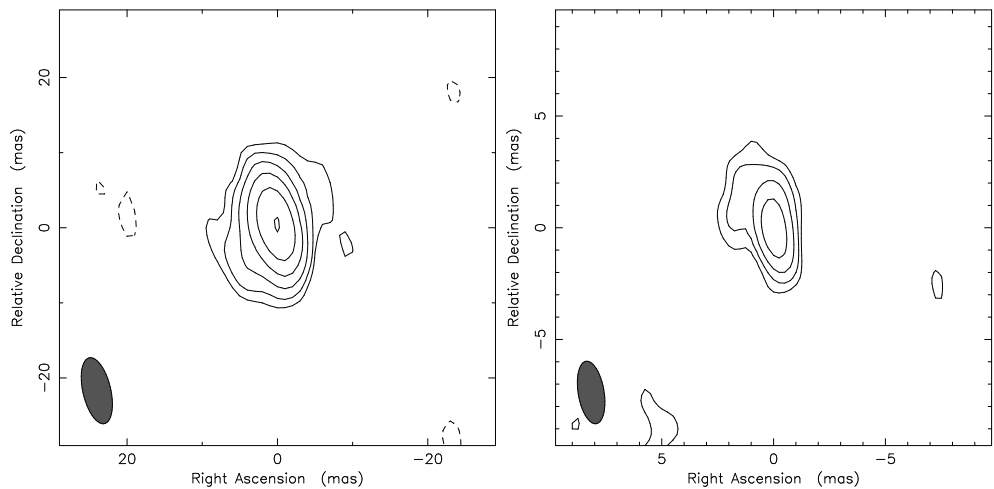

\centering
\includegraphics[bb=68 169 522 626, height=65mm, angle=270, clip=]{J1429-lband.ps}
\includegraphics[bb=68 169 522 626, height=65mm, angle=270, clip=]{J1429-cband.ps}
\caption{EVN images of J1429+5447 at 1.6~GHz (left) and 5~GHz (right). In the 1.6-GHz image, the lowest contours are drawn at $\pm70$~$\mu$Jy/beam. The peak brightness is 2.32~mJy/beam. The Gaussian restoring beam is 9.0~mas$\times$3.7~mas with major axis position angle $13^{\rm o}$. In the 5-GHz image, the lowest contours are drawn at $\pm50$~$\mu$Jy/beam. The peak brightness is 0.67~mJy/beam. The Gaussian restoring beam is 2.8~mas$\times$1.2~mas with major axis position angle $9^{\rm o}$. In both images, the positive contour levels increase by a factor of 2. The restoring beams in full width at half maximum (FWHM) are indicated with ellipses in the lower-left corners. The images are centered on the 5-GHz brightness peak of which the phase-referenced absolute equatorial coordinates are right ascension $14^{\rm h}29^{\rm m}52.17629^{\rm s}$ and declination $54^{\rm o}47^{\prime}17.6309^{\prime\prime}$ (J2000), with the accuracy of 0.4~mas.} 
\label{images}
\end{figure}

The US National Radio Astronomy Observatory (NRAO) Astronomical Image Processing System (AIPS) was used for the data calibration in a standard way. The calibrated data were then exported to the Caltech Difmap package for imaging. (See e.g. \cite{Frey08} for the details of a similar data reduction and references.) The naturally-weighted images at 1.6~GHz and 5~GHz (Fig.~\ref{images}) were made after several cycles of CLEANing in Difmap. No self-calibration was applied. 

\section{Results and discussion}

There is a single dominant radio feature detected in the quasar J1429+5447 at both 1.6 and 5~GHz (Fig.~\ref{images}). The images show a somewhat resolved mas-scale structure. Difmap was used to fit circular Gaussian brightness distribution model components to the interferometric visibility data at both frequencies. The 5-GHz data (Fig.~\ref{images}, right) are well described by a component with 0.99~mJy flux density and 0.67~mas diameter (FWHM). These imply the rest-frame brightness temperature $T_{\rm B}= 7.7 \times 10^{8}$~K, which confirms the AGN origin of the quasar's radio emission. At 1.6~GHz, the best-fit model is composed of two circular Gaussians. The main component has 3.03~mJy flux density and 2.63~mas diameter. Another component at 6.37~mas to the south-east (0.27~mJy, 1.29~mas) describes the weak extension seen also in the image (Fig.~\ref{images}, left). Considering the uncertainties, the sum of the flux densities in the VLBI components (3.30~mJy) is consistent with the FIRST value (2.95~mJy). We therefore see the entire L-band radio emission of J1429+5447 originating from a $\sim$10-mas region, corresponding to the linear size less than 60~pc. 

The two-point spectral index for the dominant component of the source is steep, $\alpha$=$-1.0$. The total rest-frame 5-GHz monochromatic luminosity of J1429+5447 is $4.5 \times 10^{26}$~W~Hz$^{-1}$, comparable to other high-redshift sources (e.g. \cite{Frey11}).

We can conclude that the mas-scale radio structure of the highest-redshift radio quasar known to date, J1429+5447 ($z$=6.21), is quite similar to what we have seen in the other two $z$$\sim$6 quasars (J0836+0054 and J1427+3312). The $T_{\rm B}$$\sim$$10^{9}$~K brightness temperature suggests that relativistic beaming does not play a major role in the appearance of the source. The steep spectrum between the observed 1.6 and 5~GHz frequencies (12 and 36~GHz in the rest frame of the quasar) is consistent with the assumption that we see the compact ``hot spots'' confined within a region of $<100$~pc in a  young GPS source at an early cosmological epoch.

\end{document}